\documentclass[aps,prb,twocolumn,superscriptaddress,nofootinbib]{revtex4-2}
\usepackage{amssymb,graphicx,color}
\usepackage[intlimits]{amsmath}
\usepackage[english]{babel}
\usepackage[colorlinks]{hyperref}
\hypersetup{
colorlinks=blue,
linkcolor=blue,
filecolor=blue,
urlcolor=blue,
citecolor=blue
}
\pdfoutput=1
\graphicspath{{figures/}} 
\usepackage[normalem]{ulem}
\usepackage{comment}
\usepackage{ragged2e}
\usepackage{enumitem}
\usepackage{amsmath}

\begin{document}

\title{Quantum transport under oscillatory drive with disordered amplitude}

\author{Vatsana Tiwari}
\affiliation{Department of Physics, Indian Institute of Science Education and Research, Bhopal 462066, India}

\author{Sushanta Dattagupta}
\affiliation{Sister Nivedita University, New Town, Kolkata 700156, India}

\author{Devendra Singh Bhakuni}
\affiliation{The Abdus Salam International Centre for Theoretical Physics (ICTP), Strada Costiera 11, 34151 Trieste, Italy}
\author{Auditya Sharma}
\email{auditya@iiserb.ac.in}
\affiliation{Department of Physics, Indian Institute of Science Education and Research, Bhopal 462066, India}

\begin{abstract}

We investigate the dynamics of non-interacting particles in a
one-dimensional tight-binding chain in the presence of an electric
field with random amplitude drawn from a Gaussian distribution, and
explicitly focus on the nature of quantum transport. We derive an
exact expression for the probability propagator and the mean-squared
displacement in the clean limit and generalize it for the disordered
case using the Liouville operator method. Our analysis reveals that in
the presence a random static field, the system follows diffusive
transport; however, an increase in the field strength causes a
suppression in the transport and thus asymptotically leads towards
localization. We further extend the analysis for a time-dependent
disordered electric field and show that the dynamics of
mean-squared-displacement deviates from the parabolic path as the
field strength increases, unlike the clean limit where ballistic
transport occurs.
\end{abstract}

\maketitle

\section{Introduction}
Quantum transport has emerged as a topic of intense interest as it
offers essential insights into the realization of novel
non-equilibrium quantum phases
~\cite{dittrich1998quantum,agarwal2015anomalous,yevgeny2017transport,kloss2019spin,
  Jepsen2020spin,haldar2021dynamical,joshi2022observing,zisling2022transport,Claeys2022absence,bhakuni2024noise,
  dhawan2024anomalous,Gamba2024hyperballistic}. This field is particularly significant from
the perspective of quantum
engineering~\cite{polkovnikov2011nonequilibrium,
Barthelemy2013quantum,Eisert2015,maier2019environment,sarkar2022emergence}.
The transport features of quantum-mechanical systems can be
categorized into two main classes: \textit{ballistic} transport--
corresponding to free particles and integrable
systems~\cite{richter2000semiclassical,Zotos2002ballistic,sirker2006spin,
  sirker2011conservation,theory2016kolasiifmmode,ballistic2019ljubotina}, and
\textit{diffusive} transport-- corresponding to generically
interacting quantum
systems~\cite{espostio2005emergence,sirker2006spin,
  diffusion2009sirker,Derrico2013quantum,
  agarwal2015anomalous,michailidis2024corrections}. Besides these,
examples of systems displaying anomalous transport signatures are also
known; examples include disordered many-body systems, systems with
multifractal eigenstates and long-range interacting systems
~\cite{ mirlin1996transition,Lima2004finite,agarwal2015anomalous,yevgeny2017transport,
purkayastha2018anomalous,Hanggi2020many,joshi2022observing,
Claeys2022absence,dhawan2024anomalous}. Furthermore,
it is also possible for transport to be entirely
absent~\cite{anderson1958absence,
  aubry1980analyticity,hone1993locally,agarwal2015anomalous}. Systems
that exhibit the phenomenon of Anderson localization, where the
presence of random disorder leads to the suppression of transport due
to the localization of the single-particle
states~\cite{anderson1958absence,
abrahams1979scaling,evers2008anderson} are a paradigmatic
example. Alongside those with random disorder, systems with
quasiperiodic disorder are also known to feature localization and,
hence, a suppression of transport~\cite{aubry1980analyticity,
  Varma2017fractality, purkayastha2018anomalous,
  Ng2007wavepacket,Li2023observation}.

Single-particle localization can also be realized in disorder-free
systems subjected to a suitable static or time-periodic uniform
electric field~\cite{Ben1996bloch,
  niu1996atomic,lyssenko1997direct,waschke1993coherent}. The addition
of a static electric field and a time-periodic or quasiperiodic
electric field drive in a one-dimensional system leads to
Wannier-Stark
localization~\cite{zener1934theory,wannier1960wave,Krieger1986time},
and dynamical localization respectively~\cite{
  gong2005dynamics,LENZ200387,dunlap1986dynamic, grosmann1991coherent,
  van2019bloch,david2017absence,dunlap1986dynamic,
  tiwari2024dynamical}. Additionally, the intricate interplay of ac
and dc electric field leads to the emergence of super Bloch
oscillations~\cite{caetano2011wave,kudo2011theoretical,Longhi2012correlated,bhakuni2018characteristic},
where it is possible to preserve localization by appropriately tuned
drive parameters.  Furthermore, the inevitable coupling of
environmental fluctuations is known to induce anomalous dynamical
features~\cite{dephasing2013mendoza,noise2017gopalakrishnan,Taylor2021subdiffusion,Landi2022nonequilibrium,
  talia2022logarithmic,Dattagupta2022imprint,Davis2023probing,bhakuni2023noise,sarkar2024impact}
in an isolated system. Therefore, it is essential to investigate the
effect of a disordered potential on electric field induced phenomena
in isolated systems. Studies reveal that while Bloch oscillations
remain stable in the presence of slowly varying spatial or temporal
noise~\cite{dom2003bloch,de2005bloch}, in contrast, rapidly
fluctuating temporal noise incoherently destroys the Wannier-Stark
localization leading to ballistic
transport~\cite{Bhakuni2019effect,bandyopadhyay2020effect}. The
addition of time-fluctuating noise to a periodically driven system
preserves dynamical localization for small times, while leading to
delocalization in the long time
limit~\cite{tiwari2022noise}. Moreover, the interplay between two
localization mechanisms—electric field periodic driving and spatial
disorder, results in stable Anderson
localization~\cite{martinez2006delocalization} if the frequency of the
drive is sufficiently high.

In this work, we study a quantum system subjected to a disordered
electric field drive. To model the disordered electric field, we
consider the amplitude of the electric field to be a random variable
drawn from a Gaussian distribution. Specifically, we study a
one-dimensional tight-binding Hamiltonian subjected to a static or
time-periodic electric field with the random amplitude drawn from a
Gaussian distribution~\cite{barry2004digital}.  The one-dimensional
tight-binding chain provides an analytically solvable platform in
which some interesting questions may be answered. We treat the
amplitude of the disordered electric field as a stochastic term in the
Hamiltonian and follow the Liouville space
formalism~\cite{Haken1972,Haken1973,Mott1961the,kubo1963stochastic,
  mukamel1988statistical,kraus1991exact,das2022quantum}.  In this
method, we evaluate the average probability propagator by computing
the first and second order cumulants~\cite{kubo1963stochastic} and
averaging over the ensemble of the stochastic process. Using the
cumulant expansion theorem~\cite{dattagupta2012relaxation}, we compute
the cumulants upto second order where we perform a single average over
an exponential series that incorporates the term accounting for the
Gaussian stochastic process.

The main findings of our work are as follows. First we derive a
general expression for the probability propagator in Liouville space,
and recover the well-known analytical results for mean squared
displacement in the limits of static and oscillatory clean electric
field. In the presence of a random static field, the mean-squared
displacement shows linear growth with time suggesting diffusive
transport with a diffusion coefficient that varies inversely with the
strength of the field. On the other hand, the presence of an
oscillatory field with disordered amplitude suppresses the free
particle like ballistic transport, asymptotically leading towards
localization in the presence of very high amplitude.

We have organized our article as follows. In Section~\eqref{Model
  hamiltonian}, we introduce the model Hamiltonian along with the
unitary transformation and rewrite the Hamiltonian in concise form. In
Section~\eqref{Probability Propagator}, we derive the expression for
the probability propagator $P_{m}\left(t\right)$. We discuss the clean limit in Section \eqref{Clean Limit}, while
Sections \eqref{Random Static Field} and \eqref{Oscillatory Field} are
focused on the case of random static field and oscillatory field
respectively. We summarize our main findings in Section \eqref{Summary
  and Conclusion}. The details of the first order and second order
cumulant calculations are shown in Appendix
\ref{A:Effective_Hamiltonian}.

\section{MODEL HAMILTONIAN}\label{Model hamiltonian}
We consider a one dimensional tight-binding fermionic system under the
influence of a time-dependent sinusoidal electric field. The model
Hamiltonian can be written as:
\begin{eqnarray}
\label{H1}
\hspace{-8ex}H&=& -\frac{\Delta}{2}\left(\sum_{n=-\infty}^{n=\infty}|n\rangle \langle n+1|+h.c.\right)+H_{0}(t), \\
H_{0}(t)&=& \mathcal{F}(t)\sum_{n=-\infty}^{n=\infty}n|n\rangle \langle n|, 
\end{eqnarray}
where $|n\rangle$ is a Wannier state localized at site $n$ and
$\Delta$ is the hopping strength. $\mathcal{F}\left(t\right)$ is the
time-periodic electric field: $\mathcal{F}\left(t\right)= A+B
\cos\left(\omega t\right)$.  We define the following unitary
operators~\cite{hartmann2004dynamics,Bhakuni2019effect}:
\begin{eqnarray}
\label{eqn:eq2a}
\hat{K}&=&\sum_{n=-\infty}^{n=\infty}|n\rangle \langle n+1|,  \\
\hat{K}^{\dagger}&=&\sum_{n=-\infty}^{n=\infty}|n+1\rangle \langle n|,\\
\hat{N}&=&\sum_{n=-\infty}^{\infty}n|n\rangle \langle n|.
\end{eqnarray}
It is worth emphasizing that the unitary operators are diagonal in the
Bloch basis defined as $ |k\rangle
=\frac{1}{\sqrt{2\pi}}\sum_{n}e^{-ink}|n\rangle $. Hence, these
operators can be expressed simply as:
\begin{eqnarray}
\label{K}
\hat{K}|k\rangle &=& e^{-ik}|k\rangle, \quad \hat{K}^{\dagger}|k\rangle = e^{ik}|k\rangle.
\end{eqnarray}
They follow the commutation relations :
\begin{eqnarray}
[\hat{K},\hat{N}]=\hat{K}, [\hat{K}^{\dagger}, N]=-\hat{K}^{\dagger}, [\hat{K}, \hat{K}^{\dagger}]=0.
\end{eqnarray}
Thus, our Hamiltonian (Eq.~\ref{H1}) can be rewritten in terms of the
unitary operators as:
\begin{eqnarray}
\label{H2}
H &=& V+H_{0}(t), \nonumber\\
V &=& \frac{-\Delta}{2}\left(\hat{K}+\hat{K}^{\dagger}\right), \quad H_{0}(t)= \mathcal{F}(t)\hat{N}.
\end{eqnarray}

\section{Probability Propagator} \label{Probability Propagator}    
To understand the transport dynamics of the randomly driven
non-interacting system, we would like to obtain an expression for the
probability propagator. The probability propagator $P_{m}(t)$ is
defined as the probability of finding the particle at site $m$ given
that the particle was initially at the origin of the chain:
\begin{eqnarray}
\label{Eq9}
P_{m}(t)&=&\langle m|\rho(t)|m\rangle ,
\end{eqnarray}
where $\rho(t)$ is the density operator. The time-evolution of the
density operator is given by the equation of motion
\begin{eqnarray}
\frac{d\rho}{dt}=-i[H, \rho].
\end{eqnarray}
Following the transformation
$$\tilde{\rho}(t)=e^{i\int_{0}^{t}H_{0}(t^{\prime})dt^{\prime}}\rho(t)e^{-i\int_{0}^{t}H_{0}(t^{\prime})dt^{\prime}}, $$
the dynamics of $\tilde{\rho}$ can be studied as
\begin{eqnarray}
\frac{d\tilde{\rho}(t)}{dt}=-i[V(t),\tilde{\rho}(t)].
\end{eqnarray}
Hence, the time-evolution of the density operator in the interaction
picture is governed by the effective Hamiltonian $V(t)$ which is
defined (in the interaction
picture~\cite{Bhakuni2019effect,tiwari2022noise}) as:
\begin{eqnarray}
\label{effV}
V(t)&=& e^{\left(i\int_{0}^{t}dt^{\prime}H_{0}(t^{\prime})\right)}Ve^{-\left(i\int_{0}^{t}dt^{\prime}H_{0}(t^{\prime})\right)} \nonumber\\
&=& -\frac{\Delta}{2}\left(\hat{K}e^{-i\eta(t)}+\hat{K}^{\dagger}e^{i\eta(t)}\right),\quad  \eta(t)=\int_{0}^{t}\mathcal{F}(t^{\prime})dt^{\prime}, \nonumber\\
\end{eqnarray}
where the phase factor $\eta(t)$ in Eq.~\eqref{effV} incorporates the
time-dependence of the Hamiltonian.  While the probability propagator
for the clean version of this model has been
calculated~\cite{Bhakuni2019effect, tiwari2022noise}, the model
studied here contains disorder. To generalize the computation for the
disordered case, we compute the matrix elements of the super-operator
$\exp(-i\mathcal{L}t)$ using the
definition~\cite{Mott1961the,das2022quantum}:
\begin{eqnarray}
\label{L_mn}
(mn|e^{-i\mathbf{\mathcal{L}}(t)}|m^{\prime}n^{\prime})=\langle m|e^{-iHt}|m^{\prime}\rangle \langle n^{\prime}|e^{iHt}|n\rangle,
\end{eqnarray}
where $|mn)$ employs the usual two-index round bracket
structure~\cite{Mott1961the} and denotes a `state' in Liouville space:
\begin{eqnarray}
|m n)&=&|m\rangle\langle n|.
\end{eqnarray}
Thus, we get the following form for the probability propagator
(Eq.~\eqref{Eq9}):
\begin{eqnarray}
\label{eqP}
P_{m}(t) = \langle m|U(t)|0\rangle \langle 0|U^{\dagger}(t)|m\rangle=\langle m|\mathbf{\mathcal{U}}(t)\rho(0)|m\rangle,
\end{eqnarray}
where $|0\rangle\langle 0|\equiv |00)$ is an ordinary operator in the
usual Hilbert space~\cite{dattagupta2012relaxation}, and $\rho(t)=
U(t)\rho(0)U^{\dagger}(t) = \mathbf{\mathcal{U}}(t)\rho(0)$. The
time-evolution operator $U(t)$ in the Schr\"{o}dinger picture is
expressed as $U(t)=\exp\left(-i\int_{0}^{t}dt^{\prime} V(t)\right)$,
while $ \mathbf{\mathcal{U}}(t)$ is the time-evolution operator in
Liouville space~\cite{dattagupta2012relaxation,das2022quantum}. Hence
\begin{eqnarray}
\label{L_Pt}
  P_{m}(t)&=& (mm|\mathbf{\mathcal{U}}(t)|00)\nonumber\\
  &=&\langle m|[e^{(-i\int_{0}^{t}dt^{\prime}\mathbf{\mathcal{V}}(t^{\prime})}]|0\rangle \langle 0|m\rangle,
\end{eqnarray}
where $\mathbf{\mathcal{V}}(t)$ is the Liouville operator associated with
$V(t)$(Eq.~\ref{effV}).

Since $V(t)$ is diagonal in the Bloch representation, we
can derive a simplified expression for $P_{m}(t)$ from Eq.~\ref{eqP}:
\begin{eqnarray}
\label{Pmt}
P_{m}(t)&=&\left(\frac{1}{2\pi}\right)^{2}\int_{-\pi}^{\pi}dk\int_{-\pi}^{\pi}dk^{\prime}\exp\left(im(k-k^{\prime})\right)\nonumber\\
&&\times(kk^{\prime}|\exp^{\left(-i\int_{0}^{t}dt^{\prime}\mathbf{\mathcal{V}}(t^{\prime})\right)}|kk^{\prime}).
\end{eqnarray}
From Eq.~\eqref{K}, and properties of the matrix elements of Liouville
operator $\mathcal{V}(t)$ (Eq.~\eqref{L_mn},~\eqref{Pmt})~\cite{das2022quantum},
\begin{eqnarray}
\label{eq13}
P_{m}(t)&=&\left(\frac{1}{2\pi}\right)^{2}\int_{-\pi}^{\pi}dk\int_{-\pi}^{\pi}dk^{\prime} \exp(im(k-k^{\prime}))\times\nonumber\\
&&\exp\left[i\frac{\Delta}{2}\left(z(t)e^{ik}-z(t)e^{ik^{\prime}}+h.c.\right)\right], 
\end{eqnarray}
where $
z(t)=\int_{0}^{t}dt^{\prime}\exp(-i\eta(t^{\prime}))=x(t)+iy(t) $. We
can again simplify Eq.~\eqref{eq13} to write the probability
propagator as
\begin{widetext}
\begin{eqnarray}
\label{P1}
P_{m}(t)&=&\left(\frac{1}{2\pi}\right)^{2}\int_{-\pi}^{\pi}dk\int_{-\pi}^{\pi}dk^{\prime} \exp(im(k-k^{\prime}))
\exp\biggl( i\Delta\bigl\lbrace\left(\cos k-\cos k^{\prime}\right)x(t)-\left(\sin k-\sin k^{\prime}\right)y(t)\bigr\rbrace\biggr).
\end{eqnarray}
\end{widetext}
Eq.~\eqref{P1} provides a general expression for the probability
propagator valid for the study of a one-dimensional system subjected
to a clean electric field. We analyse this expression to investigate
the transport properties of a system sibsjected to a clean electric
field in the next section~\eqref{Clean Limit}.

\section{Transport in Clean Limit}\label{Clean Limit}

In this section, we study transport in the one-dimensional fermionic
chain subjected to a clean electric field
$\mathcal{F}(t)=A+B\cos\omega t$. To evaluate an expression for the
mean-squared width, we start with the Fourier transform of the
probability propagator~\eqref{P1}:
\begin{eqnarray}
\label{eqPq}
P_{q}(t)&=&\sum_{m}P_{m}(t)\exp(-iqm)\nonumber.
\end{eqnarray}
From Eq.~(\ref{eq13}-\ref{P1}), we obtain:
\begin{eqnarray}
P_{q}(t) &=&\frac{1}{2\pi}\int_{-\pi}^{\pi}dk\exp\lbrace i\Delta\left[\left(\cos k-\cos (k-q)\right)x(t)\right]\rbrace\times\nonumber\\
&&\exp\biggl( -i\Delta\left[\left(\sin k-\sin(k-q)\right)y(t)\right]\biggr).
\end{eqnarray}
Now we compute the second moment of the probability propagator
$P_{m}(t)$ (Eq.~\ref{P1}) to evaluate the mean squared
displacement(MSD) defined as $X(t)=d^{2}\langle n^{2}(t)\rangle$,
where we consider lattice spacing $d=1$. Hence, we have:
\begin{eqnarray}
\label{eq:n1}
X(t) &=&\sum_{n}P_{n}(t)n^{2}=-\frac{d^{2}}{dq^{2}}P_{q}(t)|_{q=0},\\
\label{eq:n2}
&=& \frac{\Delta^{2}}{2}\left[x^{2}(t)+y^{2}(t)\right].
\end{eqnarray}
We first analyze the MSD (Eq.~\eqref{eq:n2}) in the static case, where
the electric field is time-independent, $\mathcal{F}(t)=F$. From
Eq.~\eqref{effV}, we have:
\begin{eqnarray}
\eta(t)=Ft, \quad x(t)=\frac{\sin Ft}{F},\quad y(t)=\frac{1}{F}\left(\cos Ft-1\right).\nonumber\\
\end{eqnarray}
Hence, Eq.~\eqref{eq:n2} yields
\begin{eqnarray}
\label{stM}
X(t)=\left(\frac{\Delta^{2}}{F^{2}}\right)\left[1-\cos(Ft)\right].
\end{eqnarray}
The periodicity of Eq.~\eqref{stM} signifies the well-known existence
of Bloch oscillations with time-period $2\pi/F$ in a one-dimensional
tight-binding model subjected to a static electric
field. Eq.~\eqref{stM} also encompasses the special case of zero field
$F=0$ where the free fermionic system exhibits ballistic transport:
\begin{eqnarray}
X(t)=\frac{\Delta^{2}}{2}t^{2}=v_{qu}^{2}t^{2}, \quad v_{qu}\equiv \frac{\Delta}{\sqrt{2}}.
\end{eqnarray} 
Here $v_{qu}$ can be referred to as a `quantal' velocity.

Now, we return to our specific situation of oscillatory drive:
$\mathcal{F}(t) = F\cos \omega t$. Again from
Eqs.~\eqref{effV}-\eqref{eqPq}
\begin{eqnarray}
\label{Z_osc}
z(t)&=& \int_{0}^{t}d\tau \exp\left[-i\left(F/\omega\right)\sin \tau\right],\nonumber\\
&=& t\mathcal{J}_{0}(F/\omega)+\frac{2}{\omega}\int_{0}^{\omega t}d\tau\sum_{l=1}^{\infty}\mathcal{J}_{2l}(F/\omega)\cos(2l\tau)\nonumber\\
&&-\frac{2i}{\omega}\int_{0}^{\omega t}d\tau\sum_{l=0}^{\infty}\mathcal{J}_{2l+1}(F/\omega)\sin(2l+1)\tau.
\end{eqnarray}
In Eq.~\eqref{Z_osc}, the integral over $\tau$ is written in a
dimensionless form and in the last equation we have employed the
Bessel function of the first kind with $l=0$ contribution split-off
from the summation. It is evident from Eq.~\eqref{Z_osc} that the
second and third terms contain bounded oscillatory terms whose
magnitude cannot exceed $\pi/\omega$ and hence, the second and third
terms decay in the long time limit ($\omega t\gg 1$). Consequently, in
the long time limit, the term proportional to the zeroth order Bessel
function dominates, and $z(t)$ is real
($z(t)=x(t)$)~\cite{dunlap1986dynamic}. Consequently from
Eq.~\eqref{eq:n2},
\begin{eqnarray}
\label{MSD_os}
X(t)\approx (v_{qu}^{\text{eff}})^{2}t^{2}, \quad v_{qu}^{\text{eff}}\equiv \left[d\Delta \mathcal{J}_{0}(F/\omega)\right]/\sqrt{2} ,
\end{eqnarray} 
where $v_{qu}^{\text{eff}}$ is an effective quantal velocity written
in terms of a dressed tunneling frequency $\Delta
\mathcal{J}_{0}(F/\omega)$. Thus, the long time dynamics of MSD
behaves like that of a free electron, where the prefactor
($v_{qu}^{\text{eff}}$) oscillates with $x (=F/\omega)$. Furthermore,
the first zero of $\mathcal{J}_{0}(F/\omega)$ occurs at
$F/\omega=2.405$, at which $S(t)$ vanishes irrespective of the value
of time $t$ giving rise to \textit{dynamical localization}~\cite{dunlap1986dynamic}. The
present analysis also clarifies that `long time' really implies
$\omega t\gg 1$, or $t\gg \omega^{-1}$.

As emphasized at the outset all the results in this sec- tion are
contained in~\cite{dunlap1986dynamic}, albeit in the wave function
lan- guage in contrast to the presently employed density operator
method~\cite{Dattagupta2022}, but they are repeated here to provide a
platform on which further discussions on the disorder case will be
based. In particular, the setting in terms of the Liouville space
formalism~\cite{Mott1961the} and the stratagem for extracting the
moments from the $q$-dependent probability will come handy in the
subsequent treatment.

\begin{figure}[t]
\centering
\hspace{-3ex}
  \includegraphics[width=0.45\textwidth]{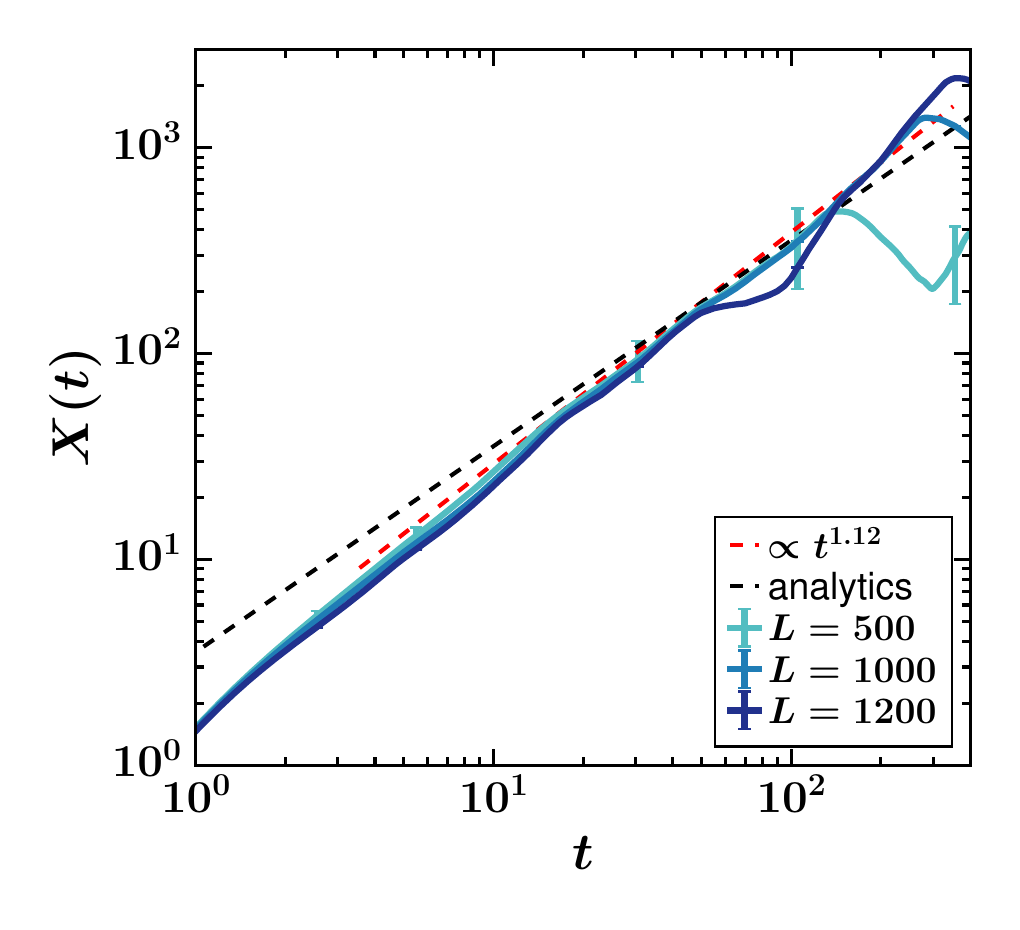}
  \caption{ Dynamics of mean squared displacement for a
    one-dimensional fermionic chain in the presence of a static
    electric field with disordered amplitude drawn from a Gaussian
    distribution with mean zero and standard deviation $F_{0}=2$. The
    data in blue shades show the numerical results and the dashed
    lines in black correspond to the analytical form. The red dashed
    line denotes the linear line ($at^{b}, b\approx 1$) fitted over
    numerical data. We have performed an average over $200$ samples
    for all the system sizes and have taken the hopping strength to be
    $\Delta=4$.}
  \label{fig:fig1a}
\end{figure}

\section{Transport in Random Static Field}\label{Random Static Field}

In the previous section, we discussed coherent dynamics in the
presence of a clean static field and a sinusoidal field. Now, we
explore the signatures of incoherence when the amplitude of the
forcing field is a quenched random variable. In this section, we
investigate the dynamics of the system in the presence of a static
electric field $\mathcal{F}(t) = F$ where we assume that the field
amplitude $F$ is extracted from a Gaussian distribution $\phi(F)$ with
mean $0$ and standard deviation $F_{0}$:
\begin{eqnarray}
\label{gauss}
\phi(F)&=& \frac{1}{2F_{0}\sqrt{\pi}}\exp(-F^{2}/4F_{0}^{2}).
\end{eqnarray}
Therefore, we need to average the probability propagator $P_{m}(t)$
(which already involves a quantum average) over a distribution of
$F$. We denote this average over the amplitude distribution by square
brackets:
\begin{eqnarray}
\left[ P_{m}(t)\right]_{\text{av}} &=& \int_{-\infty}^{\infty}dF \phi(F)P_{m}(t),
\end{eqnarray}
where $P_{m}(t)$ is obtained from Eq.\eqref{Pmt}. It is pertinent at
this stage to clarify the justification for implementing the Liouville
formalism in the present context sans which one would have to carry
out the mathematically complex task of averaging over $\phi(F)$ of the
product of separate exponential series associated with two
time-evolution operators, one to the left and the other to the right
of the initial density operator. Instead, in Liouville space, we need
to perform a single average over an exponential series, which is aided
by the application of the cumulant expansion
theorem~\cite{Mott1961the,dattagupta2012relaxation}, that yields
\begin{widetext}
\begin{eqnarray}
\label{Pm_avg}
\left[ P_{m}(t)\right]_{\text{av}} &=& \left(\frac{1}{2\pi}\right)^{2}\int_{-\pi}^{\pi}dk\int_{-\pi}^{\pi}dk^{\prime}\exp\left[im(k-k^{\prime})\right](K_{1}(t)+K_{2}(t)+K_{3}(t)),\\
\end{eqnarray}
\end{widetext}
where
\begin{eqnarray}
\label{Eq32}
K_{1}(t)&=& (kk^{\prime}|\exp\left\lbrace -i\int_{0}^{t}dt^{\prime}\left[ \hat{\mathbf{\mathcal{V}}}(t^{\prime})\right]_{\text{av}} \right\rbrace|kk^{\prime}),\\
K_{2}(t)&=& (kk^{\prime}|\left[\int_{0}^{t}dt^{\prime} \hat{\mathbf{\mathcal{V}}}(t^{\prime})\right]^{2}|kk^{\prime}),\\
K_{3}(t)&=& (kk^{\prime}|\exp\left\lbrace -\int_{0}^{t}dt^{\prime}\int_{0}^{t^{\prime}}dt^{\prime\prime}\left[ \hat{\mathbf{\mathcal{V}}}(t^{\prime})\hat{\mathbf{\mathcal{V}}}(t^{\prime\prime})\right]_{\text{av}}\right\rbrace|kk^{\prime}).\nonumber\\
\end{eqnarray}

In the appendix, we show in detail the evaluation of the first and
second order cumulants (Eqn.~\ref{Pm_avg} and Eqn.~\ref{Eq32}) in the
long-time limit ($t\rightarrow \infty$). Here we simply collect simplified expressions. For the first order cumulant, we have:
\begin{eqnarray}
\label{Eq33}
K_{1}(t)&=& \exp\left\lbrace i\frac{\Delta\sqrt{\pi}}{4F_{0}} [\cos k-\cos k^{\prime}]\right\rbrace ,
\end{eqnarray}
while the second order cumulants are given by:
\begin{eqnarray}
\label{Eq34}
K_{2}(t)&=&\pi(\Delta^{2}/F_{0}^{2})\left\lbrace\cos(k)-\cos(k^{\prime})\right\rbrace ^{2},
\end{eqnarray}
and
\begin{widetext}
\begin{eqnarray}
\label{Eq35}
K_{3}(t)&=& \exp\left\lbrace -\frac{\Delta^{2}}{4F_{0}^{2}}\left(\cos 2k+\cos 2k^{\prime}-2\cos(k+k^{\prime})\right) -\frac{\pi\Delta^{2}}{F_{0}^{2}}\left\lbrace \cos(k)-\cos(k^{\prime})\right\rbrace -2\sqrt{\pi}\frac{\Delta^{2}t}{F_{0}}\sin^{2}(k-k^{\prime})/2\right\rbrace.
\end{eqnarray}
\end{widetext}
A huge simplification is now possible since it so happens that all the
terms except the third term of Eq.~\eqref{Eq35} can be ignored.  An
elaborate discussion of the reasoning behind ignoring these terms is
contained in the Appendix (\ref{A:Effective_Hamiltonian}). After this simplification, we obtain a
compact expression for the probability propagator:
\begin{eqnarray}
\label{Eq36}
\left[ P_{m}^{2}(t)\right]_{\text{av}} &=&\left(\frac{1}{2\pi}\right)^{2}\int_{-\pi}^{\pi}dk \int_{-\pi}^{\pi}dk^{\prime}
\exp(im(k-k^{\prime}))\nonumber\\
&&\times \exp\left\lbrace{-2\sqrt{\pi}(\Delta^{2}/F_{0})t \sin^{2}(k-k^{\prime})/2}\right\rbrace.\nonumber\\
\end{eqnarray}
Next, we compute the Fourier transform of Eq.~\eqref{Eq36}, and
evaluate the mean squared width. We have:
\begin{eqnarray}
\label{Eq:msd_r}
X(t)&=& -\frac{d^{2}}{dq^{2}}\left[ P^{(2)}(q,t)\right]_{\text{av}}=\sqrt{\pi}\left(\Delta^{2}/F_{0}\right)t.\nonumber\\
\end{eqnarray}

In order to analyze the difference between the transport behavior of
the system in the presence of a clean static field~\eqref{stM} and
random static field~\eqref{Eq:msd_r}, we compare the dynamics of the
mean squared displacement. Eq.~\eqref{stM} shows that the the uniform
static field leads to Bloch oscillations, where the amplitude of the
oscillations decreases with an increase in the field strength $F$. In
contrast, in the presence of a random static field, the dynamics of
mean squared displacement (MSD) exhibits power-law growth,
$X(t)\propto t^{\alpha}$, where $\alpha$ corresponds to diffusive
transport as shown in Fig.~\eqref{fig:fig1a} for $F_{0}=2$ of the
Gaussian distribution~\eqref{gauss}. It is worth mentioning that our
analytical curve starts to agree with our numerical data when $F_{0}t$
is sufficiently large so that the diffusive limit holds. However, in
the long time limit, the agreement between the two breaks down
(Fig.~\eqref{fig:fig1a}) and the numerical data shows saturating
behaviour.

The result in Eq.~\eqref{Eq:msd_r} is also reminiscent of the classical
problem of a random walk on a chain in which the walker can jump to
the right or to the left with equal probability at a mean rate
$\nu$. In the absence of a bias, the MSD has a diffusive behavior with
\begin{eqnarray}
\label{diff}
X(t)&=& 2Dt=2\nu d^{2}t,
\end{eqnarray}
where $D$ is the diffusion constant, and $d$ is a lattice constant. The
symmetric Gaussian distribution and hence zero mean bias facilitates
the comparison between Eq.~\eqref{Eq:msd_r} and ~\eqref{diff} which leads
to the mean jump rate $\nu$ given by
\begin{eqnarray}
\label{nu}
\nu &=& \sqrt{\pi}\Delta^{2}/2F_{0}.
\end{eqnarray}
Thus, the jumps become less frequent for a modest tunneling frequency
$\Delta$ and large width $F_{0}$ of the underlying Gaussian
distribution of the fields - a signature of disorder in the amplitude
leading to localization. Interestingly, the same diffusive behaviour
ensues in a strongly dissipative model of quantum transport in a
tight-binding lattice when the field $F$ is not an external one but an
internal boson field in an open system when the tight binding chain is
coupled to a bosonic heat bath~\cite{BM2021dissipative}. Naturally,
the diffusion coefficient in that case depends on the temperature and
the microscopic details of the bath such as the cut-off frequency of
an underlying Ohmic dissipation scenario of the bosonic
excitations. Because of these reasons, we would like to view the
result for the MSD in the disordered electric field driven
tight-binding chain as a signature of incoherent transport.

\vspace{2ex}

\section{Transport in an oscillatory field with random amplitude}\label{Oscillatory Field}
\begin{figure}[t]
\centering
\hspace{-3ex}
  \includegraphics[width=0.5\textwidth]{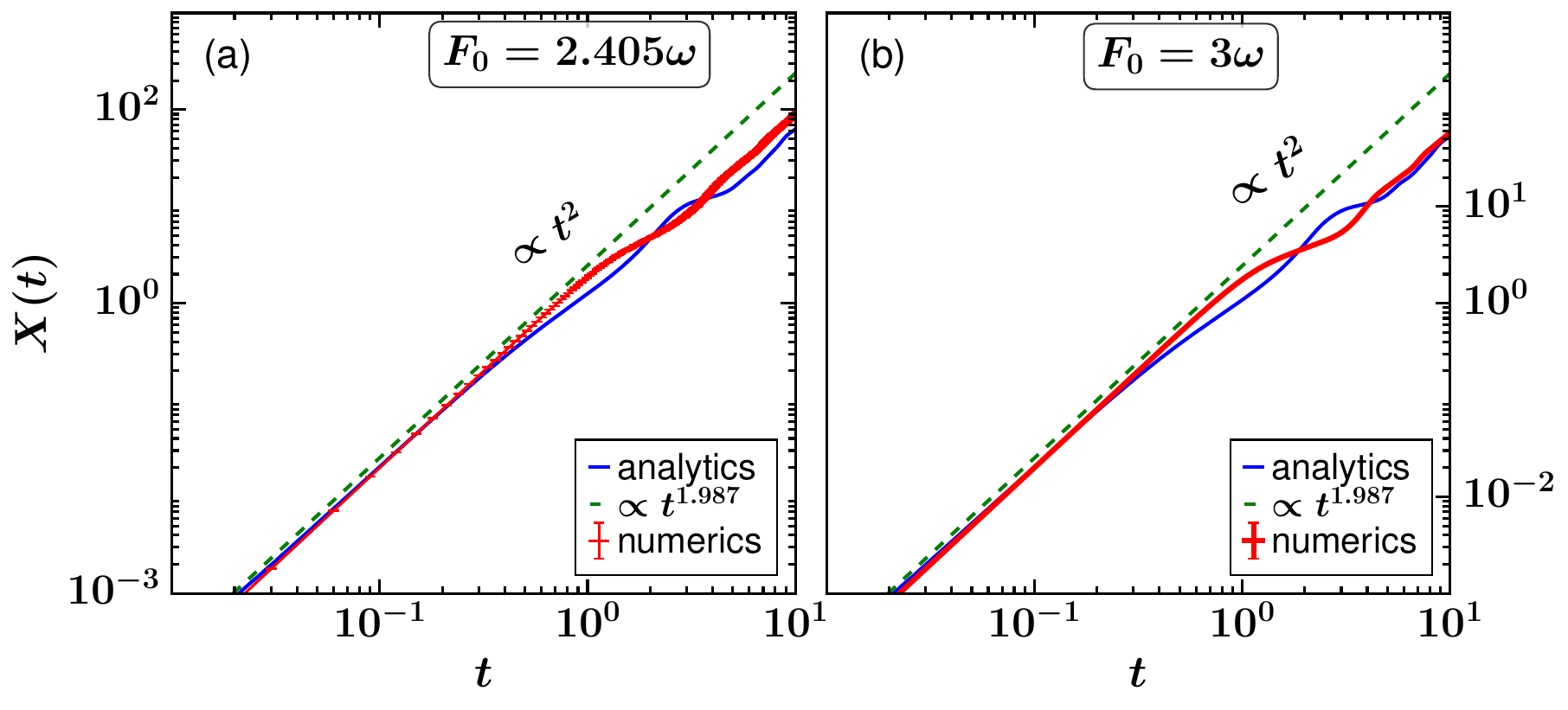}
 	\caption{Dynamics of mean squared width for periodically
          driven system with disordered amplitude drawn from a
          Gaussian distribution with mean zero and standard deviation
          $F_0$. (a) Dynamical localization (DL) point for the clean
          limit with $F_{0}=2.405\omega$, (b) Away from dynamical
          localization point (ADL) with $F_{0}=3\omega $. The green
          dashed line shows power-law transport given by $X(t)\propto t^{2}$. The other
          parameters are $\Delta=4$, $\omega=1$, and system size
          $L=200$.}
  \label{fig:fig2}
\end{figure}
In this section, we study the transport in a one dimensional fermionic
chain subjected to a periodically driven electric field with a
disordered amplitude. We set $\mathcal{F}(t)=F\cos(\omega t)$ where
the drive comes with a random amplitude drawn from a Gaussian
distribution (Eq.\ref{H1},Eq.\ref{gauss}). Following the analysis of
Sec.~\eqref{Random Static Field}, we restrict the discussion to the
long time limit, and hence focus on the second order cumulant. We
compute the correlation function, the central quantity of interest as:
\begin{eqnarray}
\label{V_osc}
\left[ \mathbf{\mathcal{V}}(t^{\prime})\mathbf{\mathcal{V}}(t^{\prime\prime})\right]_{\text{av}} &=& \frac{\Delta^{2}}{4}\left[ \lbrace(\hat{\mathbf{\mathcal{K}})}^{2}e^{-iF(\xi(t^{\prime})+\xi(t^{\prime\prime}))}+\right.\nonumber\\
&&\left.(\hat{\mathbf{\mathcal{K}}}^{\dagger})^{2}e^{iF(\xi(t^{\prime})+\xi(t^{\prime\prime}))}\right.\nonumber\\
&&\left. +\hat{\mathbf{\mathcal{K}}}\hat{\mathbf{\mathcal{K}}}^{\dagger}[e^{iF(\xi(t^{\prime})-\xi(t^{\prime\prime}))}+c.c.]\rbrace\right]_{av} ,\nonumber\\
\end{eqnarray}
where 
\begin{eqnarray}
\label{xi}
\xi(t)&=&\frac{1}{\omega}\sin(\omega t).
\end{eqnarray}
Convinced by earlier analysis in the static case, we will only
consider the cross terms, i.e.,
\begin{eqnarray}
\left[ \mathbf{\mathcal{V}}(t^{\prime})\mathbf{\mathcal{V}}(t^{\prime\prime})\right]_{av} &\rightarrow & \frac{\Delta^{2}}{4}{\hat{\mathbf{\mathcal{K}}}\hat{\mathbf{\mathcal{K}}^{\dagger}}\left[ [e^{iF(\xi(t^{\prime})-\xi(t^{\prime\prime})}+c.c.]\right]_{av} }.\nonumber\\
\end{eqnarray}
Averaging over the random distribution of $F$ leads to the following
form of the correlator:
\begin{widetext}
\begin{eqnarray}
\left[
  \mathbf{\mathcal{V}}(t^{\prime})\mathbf{\mathcal{V}}(t^{\prime\prime})\right]_{av}
&\rightarrow &
\frac{\Delta^{2}}{4}\int_{0}^{t}dt^{\prime}\int_{0}^{t^{\prime}}dt^{\prime\prime}\exp\biggl(-(F_{0})^{2}\left(\xi
  \left(t^{\prime}\right)-\xi
  \left(t^{\prime\prime}\right)\right)^{2}\biggr)\mathbf{\hat{\mathcal{K}}}\mathbf{\hat{\mathcal{K}^{\dagger}}}.
\end{eqnarray}
Next, in the Liouville space, we can rewrite the expression as
\begin{eqnarray}
(kk^{\prime}|\int_{0}^{t}dt^{\prime}\int_{0}^{t^{\prime}}dt^{\prime\prime}\left[\mathbf{\mathcal{V}}(t^{\prime})\mathbf{\mathcal{V}}(t^{\prime\prime})\right]_{\text{av}}|kk^{\prime})\rightarrow -\frac{\Delta^{2}}{4}\int_{0}^{t}dt^{\prime}\int_{0}^{t^{\prime}}dt^{\prime\prime}\exp\biggl(-(F_{0})^{2}(\xi(t^{\prime})-\xi(t^{\prime}))^{2}\biggr)\sin^{2}[(k-k^{\prime})/2].
\end{eqnarray}
Consequently (cf., Eq.~\eqref{eqPq}),
\begin{eqnarray}
\left[ P_{q}(t)\right]_{\text{av}} &\rightarrow & \exp\left\lbrace-\frac{\Delta^{2}}{4}\int_{0}^{t}dt^{\prime}\int_{0}^{t^{\prime}}dt^{\prime\prime}
\exp\biggl(-(F_{0})^{2}\bigl(\xi(t^{\prime})-\xi(t^{\prime\prime})\bigr)^{2}\biggr)\sin^{2}q/2\right\rbrace.\nonumber\\
\end{eqnarray}
\end{widetext}
This leads to the following expression for MSD (from
Eq.~\eqref{eq:n1}) for our system in the presence of a time-periodic
electric field with random amplitude:
\begin{eqnarray}
X(t)=\frac{\Delta^{2}}{2}\int_{0}^{t}dt^{\prime}\int_{0}^{t^{\prime}}dt^{\prime\prime}\exp\biggl(-F_{0}^{2}(\xi(t^{\prime})-\xi(t^{\prime\prime}))^{2}\biggr),\nonumber\\
\end{eqnarray}
where
\begin{eqnarray}
(\xi(t^{\prime})-\xi(t^{\prime\prime}))^{2}&=&\frac{1}{\omega^{2}}\biggl[\sin(\omega t^{\prime})-\sin(\omega t^{\prime\prime})\biggr]^{2}.
\end{eqnarray}
Thus, in the long time limit, we obtain the expression:
\begin{eqnarray}
\label{msd_f}
X(t)&=&\frac{\Delta^{2}}{2}\int_{0}^{t}dt^{\prime}I(t^{\prime}),
\end{eqnarray}
where
\begin{eqnarray}
I(t^{\prime})&=&\int_{0}^{t^{\prime}}dt^{\prime\prime}\exp\left\lbrace-\frac{F_{0}^{2}}{\omega^{2}}[\sin(\omega t^{\prime})-\sin (\omega t^{\prime\prime})]^{2}\right\rbrace.\nonumber\\
\end{eqnarray}
\begin{figure}[t]
\centering
\hspace{-3ex}
  \includegraphics[scale=0.47]{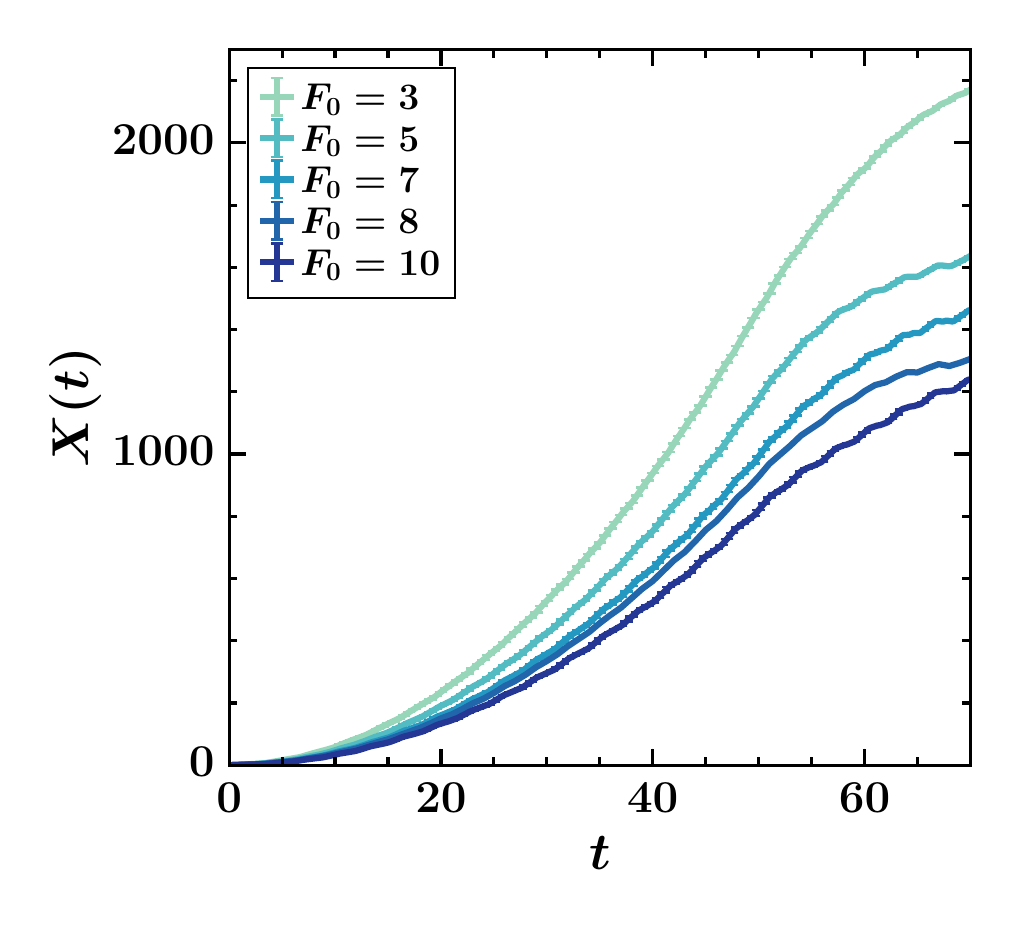}
 	\caption{Dynamics of mean-squared width for periodically driven system with random amplitude drawn from Gaussian distribution for various values of standard deviations $F_{0}$. The other parameters are $\Delta=4$, and $ L=200$.}
  \label{fig:fig3}
\end{figure}
Next, in order to make a meaningful analysis of how dynamical
localization is affected by the presence of static disorder we pull
out the time $t$ from the integral in Eq.~\eqref{msd_f} to write the
latter in a dimensionless form, thus
\begin{widetext}
\begin{eqnarray}
X(t)&=& \frac{\Delta^{2}}{2}t^{2}\left[\int_{0}^{1}d\tau \exp\biggl[-(F_{0}/\omega)^{2}\lbrace \sin(\omega t)-\sin(\omega t\tau)\rbrace^{2}\biggr]-\int_{0}^{1}d\tau \tau G(\tau)\right], \nonumber\\
\label{eq:msd_T}
\end{eqnarray}
where
\begin{eqnarray}
\label{eq:msd_G}
G(\tau)&=&\frac{d}{d\tau}\int_{0}^{\tau}d\tau^{\prime}\exp\biggl( -(F_{0}/\omega)^{2}\lbrace \sin(\omega t\tau)- \sin(\omega t\tau^{\prime})\rbrace^{2}\biggr).
\end{eqnarray}
\end{widetext}
Eq.~\eqref{msd_f} remains valid and matches with
Eq.~\eqref{Eq:msd_r} for the case of a random static field when
$\omega \rightarrow 0 $ and the upper limit of the integral is taken
to $\infty$.

\begin{figure}[t]
\centering
\hspace{-3ex}
  \includegraphics[scale=0.4]{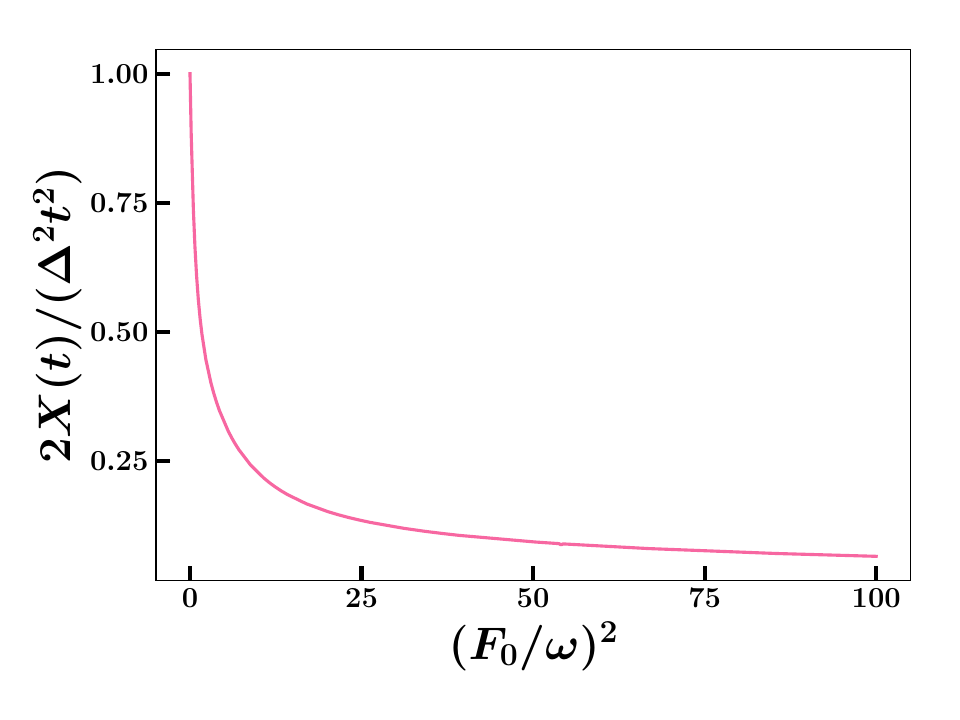}
 	\caption{$2X(t)/ (\Delta^{2}t^{2})$ versus $F_{0}/\omega $ for randomly driven system at time $t=100 $. We fix $\omega=1$ and vary $F_{0}/\omega $ to solve the integration involved in Eq.~\eqref{eq:msd_T}.}
  \label{fig:fig4}
\end{figure}

For pure oscillatory drive, MSD exhibits parabolic dependence on time
on tuning the drive parameters away from the dynamical localization
point. For random oscillatory drive, we discuss the dynamics of MSD in
Fig.~\eqref{fig:fig2} with the help of a numerical calculation of
Eq.~\eqref{eq:msd_T}. Eq.~\eqref{eq:msd_T} shows that for the
disordered case, a third frequency scale $\omega t$ appears in the
integrand of Eq.~\eqref{eq:msd_T} apart from $\Delta t$ and
$F_{0}/\omega $ which provides additional dependence on time over and
above the $t^{2}-$dependence of the pre-factor. In
Fig.~\eqref{fig:fig2}, we show the time-dependence of MSD for
$F_{0}=2.405\omega$ (dynamical localization point in clean limit), $F=
3\omega $ (away from dynamical localization point in clean limit)
which follow power-law growth, $X(t)\propto t^{2}$. 

Fig.~\eqref{fig:fig3} shows that the presence of disorder in the
amplitude of the oscillatory drive slows down the free-particle like
parabolic dependence on time. An increase in the drive-amplitude
suppresses the transport - this hints at the existence of
disorder-induced localization in the system for a sufficiently high
amplitude oscillatory drive. In order to establish the effect of a
disordered field on dynamical localization, in Fig.~\eqref{fig:fig4},
we plot $I(t)=2X(t)/(\Delta t)^{2}$ against $(F_{0}/\omega)^{2}$ for
$t=100$ in the integrand of Eq.~\eqref{eq:msd_T}. We find that the
MSD divided by the dimensionless quantity $(\Delta t)^{2}/2$ decreases
rapidly. The larger $F_{0}$ is, the faster is the fall. This behavior
is in conformity with our expectation that a large value of the
standard deviation $F_{0}$ of the probability distribution of $F$
implies stronger tendency towards localization as is clear from
Fig.~\eqref{fig:fig3}.

\section{Summary and Conclusion}\label{Summary and Conclusion}

In our study, we explore the novel dynamical characteristics of a
quantum system subjected to a disordered electric field
drive. Specifically, we study a one-dimensional tight-binding
Hamiltonian subjected to a static or time-periodic electric field with
the random amplitude drawn from a Gaussian
distribution. We begin with the derivation of an exact expression for the
probability propagator, which enables us to derive the mean-squared
displacement for both time-independent fields and time-periodic drives
in the clean limit. We further extend our analysis to understand the
disordered case with the aid of the Liouville space formalism.  With
this approach, we obtain a general expression for the probability
propagator, and subsequently an expression for the mean-squared
displacement.

In the presence of a clean electric field, we recover the well-known
results of Bloch oscillations and dynamical localization in the
time-independent and periodic cases, respectively. In the presence of
a random static field, the analytical expression for MSD shows a
linear dependence on time that indicates diffusive transport, thus
mimicking the classical problem of a random walk. However, the
increase in field strength impedes transport, and asymptotically leads
to localization. Moreover, the presence of random oscillatory drive
brings parabolic dependence of MSD on time in the long time limit,
akin to the clean, oscillatory-driven case away from the dynamical
localization points. However, a comparison between the dynamics
reveals that the disordered amplitude effectively slows down
transport, and an increase in the amplitude induces localization.

Our study finds its relevance in experimental systems where environmental fluctuations, disorder, and noise cannot be ignored. Our theoretical analysis aligns with the recent experimental and theoretical research works~\cite{environment2019Maier,Coates2021localisation,
observation2022sajjad,Davis2023probing} where the effects of environmental disorder on the dynamics of the transport of quantum systems have been explored.

\section*{Acknowledgments}
We are grateful to the High Performance Computing(HPC) facility at
IISER Bhopal, where large-scale calculations in this project were
run. V.T is grateful to DST-INSPIRE for her PhD fellowship. SD is
grateful to the Indian National Science Academy for support through
its Honorary Scientist scheme. A.S acknowledges financial support from
SERB via the grant (File Number: CRG/2019/003447), and from DST via
the DST-INSPIRE Faculty Award [DST/INSPIRE/04/2014/002461].

\newpage
\begin{widetext}
\appendix

\section{}
\section*{Probability Propagator for random static field}\label{A:Effective_Hamiltonian}

We find the following expression for probability propagator with the help of cumulant expansion theorem~\cite{dattagupta2012relaxation,Mott1961the},

\begin{eqnarray}
\label{PP}
\left[ P_{m}(t)\right]_{\text{av}} &=& \left(\frac{1}{2\pi}\right)^{2}\int_{-\pi}^{\pi}dk\int_{-\pi}^{\pi}dk^{\prime}\exp\left[im(k-k^{\prime})\right](kk^{\prime}|\exp\left\lbrace -i\int_{0}^{t}dt^{\prime}\left[\mathbf{\mathcal{V}}(t^{\prime})\right]_{\text{av}} -\left(\int_{0}^{t}dt^{\prime}\int_{0}^{t^{\prime}}dt^{\prime\prime}\left[\mathbf{\mathcal{V}}(t^{\prime})\mathbf{\mathcal{V}}(t^{\prime\prime})\right]_{\text{av}}\right)\right.\nonumber\\
&&\left.-\left(\int_{0}^{t}dt^{\prime}\left[ \mathbf{\mathcal{V}}(t^{\prime})\right]_{\text{av}}\right)^{2}\right\rbrace|kk^{\prime}).
\end{eqnarray}

\subsection{First Order Cumulant}

To analyse Eq.~\eqref{Pm_avg}, we first evaluate
$\int_{0}^{t}dt^{\prime} \left[
\mathbf{\mathcal{V}}(t^{\prime})\right]_{\text{av}} $ with the help of
Eq.~\eqref{effV}, where

\begin{eqnarray}
\biggl[ \mathbf{\mathcal{V}}(t^{\prime})\biggr]_{\text{av}}& = & \int_{-\infty}^{\infty}dF \phi(F)\biggl[\hat{\mathbf{\mathcal{K}}}e^{-iFt}+\hat{\mathbf{\mathcal{K}}}^{\dagger}e^{iFt}\biggr]_{\text{av}}.\nonumber\\
\end{eqnarray}

Hence, the first order cumulant can be expressed in compact form as

\begin{eqnarray}
\exp\left\lbrace -i\int_{0}^{t}dt^{\prime}\left[ \hat{\mathbf{\mathcal{V}}}(t^{\prime})\right]_{\text{av}}  \right\rbrace = \exp\left\lbrace i\Delta \left(\hat{\mathbf{\mathcal{K}}}+\hat{\mathbf{\mathcal{K}}}^{\dagger}\right)^{\times}\chi(t)\right\rbrace, \nonumber\\
\end{eqnarray}

where $\chi(t)$ is the integration defined as

\begin{eqnarray}
\chi(t)&=& \int_{0}^{t}dt^{\prime}\int_{-\infty}^{\infty}dF \phi(F)\exp(\pm iFt^{\prime}) \nonumber\\
& = &\int_{0}^{t}dt^{\prime}\exp\left(-F_{0}^{2}t^{\prime 2}\right)=\frac{\sqrt{\pi}}{2F_{0}}\left[1-\text{erfc}(F_{0}t)\right],\quad \left(\text{erfc}(z)= \frac{2}{\sqrt{\pi}}\int_{z}^{\infty}dz^{\prime}\exp(-z^{\prime 2})\right).\nonumber\\
\end{eqnarray}


Using the properties of the Liouville operators $\hat{\mathbf{\mathcal{K}}}$ and $\hat{\mathbf{\mathcal{K}}}^{\dagger}$ in the $k-$ representation, we get

\begin{eqnarray}
\label{A5}
(kk^{\prime}|\exp\left\lbrace -i\int_{0}^{t}dt^{\prime}\left[ \hat{\mathbf{\mathcal{V}}}(t^{\prime})\right]_{\text{av}} \right\rbrace|kk^{\prime})&=& \exp\left\lbrace i\frac{\Delta}{2} \chi(t)[\cos k-\cos k^{\prime}]\right\rbrace .\nonumber\\
\end{eqnarray}

In the long time limit $t\rightarrow \infty $, we can replace $\chi(t)$ by
\begin{eqnarray}
\chi(t=\infty)&=& \int_{0}^{\infty}d\tau \exp\left(-F_{0}^{2}\tau^{2}\right)=\frac{\sqrt{\pi}}{2F_{0}},
\end{eqnarray}

and Eq.~\eqref{A5} can be simplified as

\begin{eqnarray}
\label{A7}
(kk^{\prime}|\exp\left\lbrace -i\int_{0}^{t}dt^{\prime}\left[ \hat{\mathbf{\mathcal{V}}}(t^{\prime})\right]_{\text{av}} \right\rbrace|kk^{\prime})&=& \exp\left\lbrace i\frac{\Delta\sqrt{\pi}}{4F_{0}} [\cos k-\cos k^{\prime}]\right\rbrace .
\end{eqnarray}

\subsection{Second Order Cumulant}

\begin{eqnarray}
(kk^{\prime}|\exp\left\lbrace -\int_{0}^{t}dt^{\prime}\int_{0}^{t^{\prime}}dt^{\prime\prime}\left[ \hat{\mathbf{\mathcal{V}}}(t^{\prime})\hat{\mathbf{\mathcal{V}}}(t^{\prime\prime})\right]_{\text{av}}\right\rbrace|kk^{\prime}) &=& \exp\left\lbrace-\int_{0}^{t}dt^{\prime}\int_{0}^{t^{\prime}}dt^{\prime\prime}\biggl[ (kk^{\prime}|\hat{\mathbf{\mathcal{V}}}(t^{\prime})|kk^{\prime})(kk^{\prime}|\hat{\mathbf{\mathcal{V}}}(t^{\prime\prime})|kk^{\prime})\biggr]_{\text{av}}\right\rbrace.
\nonumber\\
\end{eqnarray}

Now,
\begin{eqnarray}
\label{eq1}
(kk^{\prime}|\hat{\mathbf{\mathcal{V}}}(t^{\prime})|kk^{\prime})(kk^{\prime}|\hat{\mathbf{\mathcal{V}}}(t^{\prime\prime})|kk^{\prime})&=&\frac{\Delta^{2}}{4}(kk^{\prime}|[\hat{\mathbf{\mathcal{K}}}e^{-iFt^{\prime}}+\hat{\mathbf{\mathcal{K}}}^{\dagger}e^{iFt^{\prime}}]|kk^{\prime})(kk^{\prime}|[\hat{\mathbf{\mathcal{K}}}e^{-iFt^{\prime\prime}}+\hat{\mathbf{\mathcal{K}}}^{\dagger}e^{iFt^{\prime\prime}}]|kk^{\prime})\nonumber\\
&=& \frac{\Delta^{2}}{4}\biggl[(e^{ik}-e^{ik^{\prime}})^{2}e^{iF(t^{\prime}+t^{\prime\prime})}+(e^{-ik}-e^{-ik^{\prime}})^{2}e^{-iF(t^{\prime}+t^{\prime\prime})}\nonumber\\
&&+2(e^{-ik}-e^{-ik^{\prime}})(e^{ik}-e^{ik^{\prime}})\cos(F(t^{\prime}-t^{\prime\prime}))\biggr]\nonumber\\
&=&\frac{\Delta^{2}}{4}\biggl[(e^{ik}-e^{ik^{\prime}})^{2}e^{iF(t^{\prime}+t^{\prime\prime})}+(e^{-ik}-e^{-ik^{\prime}})^{2}e^{-iF(t^{\prime}+t^{\prime\prime})}\nonumber\\
&&+8(\sin^{2}(k-k^{\prime})/2)\cos(F(t-t^{\prime\prime}))\biggr],
\end{eqnarray}
where we have employed the definition of matrix elements of Liouville operator~\cite{das2022quantum}:

\begin{eqnarray}
\label{liou}
(nm|\mathcal{L}|n^{\prime}m^{\prime})&=& \langle n|H|n^{\prime}\rangle \delta_{mm^{\prime}} - \langle m^{\prime}|H|m\rangle \delta_{nn^{\prime}},
\end{eqnarray}
and $\mathcal{L}$ is the Liouville operator associated with the Hamiltonian $H$.
Next, we solve the time integration involved in each term of Eq.~\eqref{eq1} one by one. First, we solve time-dependent term in the third term

\begin{eqnarray}
\label{A2}
I_{1} &=& \int_{0}^{t}dt^{\prime}\int_{0}^{t^{\prime}}dt^{\prime\prime}\cos(F(t^{\prime}-t^{\prime\prime}))= 2\int_{0}^{t}dt^{\prime}\int_{0}^{t^{\prime}}d\tau \cos(F\tau)\nonumber\nonumber\\
&=& 2\int_{0}^{t}d\tau (t-\tau)\cos(F\tau)\rightarrow 2t\chi(t)-[1-\exp(-F_{0}^{2}t^{2})]/F_{0}^{2}\equiv 2t\chi^{\prime}(t),
\end{eqnarray}

where we have performed average over $F$, and $\chi^{\prime}(t)=\chi(t)-[1-\exp(-F_{0}^{2}t^{2})]/2tF_{0}^{2}$. However, in the long-time limit, $\chi^{\prime}(t)$ can be replaced by $\chi(t)$ and eventually, by $\chi(t=\infty)$. Thus, the corresponding (dominant) term results in the simplification of third term of Eq.~\eqref{A2}
\begin{eqnarray}
\frac{\Delta^{2}}{4}\left(8\sin^{2}(k-k^{\prime})/2. I_{1}\right)=2\sqrt{\pi}(\Delta^{2}/F_{0})t.\sin^{2}(k-k^{\prime})/2.
\end{eqnarray}

Now, we solve the terms in the first and second term,

\begin{eqnarray}
\langle e^{iF(t^{\prime}+t^{\prime\prime})}\rangle &=& \langle e^{-iF(t^{\prime}+t^{\prime\prime})}\rangle =\exp[-F_{0}^{2}(t^{\prime}+t^{\prime\prime})^{2}],
\end{eqnarray}

and the double integration yields
\begin{eqnarray}
\label{eq6}
2\int_{0}^{t}dt^{\prime}\int_{0}^{t^{\prime}}dt^{\prime\prime}\exp[-F_{0}^{2}(t^{\prime}+t^{\prime\prime})^{2}]&=&2\int_{0}^{t}\int_{t^{\prime}}^{2t^{\prime}}d\tau \exp(-F_{0}^{2}\tau^{2})\nonumber\\
&= & 2 t\int_{t}^{2t}d\tau\exp(-F_{0}^{2}\tau^{2})-\int_{0}^{t}d\tau \tau[2\exp(-4F_{0}^{2}\tau^{2})-\exp(-F_{0}^{2}\tau^{2})]\nonumber\\
&=& 2t\int_{0}^{t}d\tau \exp[-F_{0}^{2}(t+\tau)^{2}]+\frac{1}{2F_{0}^{2}}[1-2\exp(-F_{0}^{2}t^{2})+\exp(-4F_{0}^{2}t^{2})].
\end{eqnarray}

In the long time limit, Eq.~\eqref{eq6} yields $1/(2F_{0}^{2})$. This term multiplies
\begin{eqnarray}
[(e^{ik}-e^{ik^{\prime}})^{2}+C.C]&=& 2[\cos 2k+\cos 2k^{\prime}-2\cos(k+k^{\prime})].
\end{eqnarray}

Hence, in the long time limit, the corresponding double-integral term leads to

\begin{eqnarray}
\label{A15}
\frac{1}{F_{0}^{2}}\left(\cos 2k+\cos 2k^{\prime}-2\cos(k+k^{\prime}\right).
\end{eqnarray}

Next, we compute the third term in Eq.~\eqref{PP} as,

\begin{eqnarray}
\label{A16}
\Biggl[\int_{0}^{t}dt^{\prime} (kk^{\prime}|\hat{\mathbf{\mathcal{V}}}(t^{\prime})|kk^{\prime})\Biggr]^{2}=\pi(\Delta^{2}/F_{0}^{2})\Biggl[\cos(k)-\cos(k^{\prime})\Biggr] ^{2}.
\end{eqnarray}

Thus, in long time limit, we recover following expression for probability propagator with the help of first and second order cumulants (Eq.~\eqref{A7},~\eqref{A15}, and ~\eqref{A16}),

\begin{eqnarray}
\label{A10}
\left[ P_{m}^{2}(t)\right]_{\text{av}} &=& \left(\frac{1}{2\pi}\right)^{2}\int_{-\pi}^{\pi}dk \int_{-\pi}^{\pi}dk^{\prime}\exp(im(k-k^{\prime}))\exp\left\lbrace i\frac{\sqrt{\pi}\Delta }{4F_{0}}\left[\cos k-\cos k^{\prime}\right]-\frac{\Delta^{2}}{4F_{0}^{2}}\left(\cos 2k+\cos 2k^{\prime}-2\cos(k+k^{\prime}\right)\right.\nonumber\\
&&\left.-2\sqrt{\pi}\frac{\Delta^{2}t}{F_{0}}\sin^{2}(k-k^{\prime})/2-\frac{\pi\Delta^{2}}{F_{0}^{2}}\left\lbrace \cos(k)-\cos(k^{\prime})\right\rbrace\right\rbrace.
\end{eqnarray}

Now, it is evident from Eq.~\eqref{A10} that as $t\rightarrow \infty $, the term proportional to $\left(\Delta t/F_{0}\right)^{2}$ dominates over terms proportional to $\left(\Delta/F_{0}\right)$ and $\left(\Delta t/F_{0}\right)^{2}$, which implies that $\left(\Delta^{2}t\right)\gg 1 $ and $\left(F_{0}t\right)\gg 1 $ . Thus,

\begin{eqnarray}
\label{P_avt}
\left[ P_{m}^{2}(t)\right]_{\text{av}} &=&\left(\frac{1}{2\pi}\right)^{2}\int_{-\pi}^{\pi}dk \int_{-\pi}^{\pi}dk^{\prime}
\exp(im(k-k^{\prime})) \exp\biggl(-2\sqrt{\pi}(\Delta^{2}/F_{0})t \sin^{2}(k-k^{\prime})/2\biggr).\nonumber\\
\end{eqnarray}

Since, our aim is to evaluate an expression for mean squared width, $\left[ m^{2}(t)\right]_{\text{av}}=-\frac{d^{2}}{dq^{2}}\left[ P^{2}(q,t)\right]_{\text{av}}|_{q=0}$, we find the Fourier transformation of Eq.~\eqref{P_avt},

\begin{eqnarray}
\left[P^{2}(q, t)\right]=\exp\biggl( -2\sqrt{\pi}\left(\Delta^{2}/F_{0}t\right)\sin^{2}q/2\biggr),
\end{eqnarray}

and compute the derivatives of $\left[ P^{(2)}(q,t)\right]_{\text{av}}$
as
\begin{eqnarray}
\frac{d}{dq}\left[ P^{2}(q,t)\right]_{\text{av}} &=&  -2\sqrt{\pi}(\Delta^{2}/F_{0})t\sin(q/2)\exp\bigl( -2\sqrt{\pi}(\Delta^{2}/F_{0})t\sin^{2}(q/2)\bigr),
\end{eqnarray}

and

\begin{eqnarray}
\frac{d^{2}}{dq^{2}}\left[ P^{2}(q,t)\right]_{\text{av}} &=&\biggl[ -2\sqrt{\pi}\left(\frac{\Delta^{2}}{F_{0}}\right)\cos(q/2)t\exp\biggl( -2\sqrt{\pi}(\Delta^{2}/F_{0})t \sin^{2}(q/2)\biggr)\biggr] +\biggl[\left(2\sqrt{\pi}(\Delta^{2}/F_{0})t\sin(q)\right)^{2}\nonumber\\
&& \times \exp\biggl( -2\sqrt{\pi}(\Delta^{2}/F_{0})t\sin^{2}(q/2)\biggr)\biggr].
\end{eqnarray}

This leads to the following expression for mean-squared width
\begin{eqnarray}
\left[ m^{2}(t)\right]_{\text{av}}&=& -\frac{d^{2}}{dq^{2}}\left[ P^{2}(q,t)\right]_{\text{av}}=\sqrt{\pi}\left(\Delta^{2}/F_{0}\right)t.
\end{eqnarray}
%

\end{widetext}
\bibliography{refv_new}
\end{document}